# Chirality and angular momentum in optical radiation

Matt M. Coles and David L. Andrews*

School of Chemistry, University of East Anglia, Norwich NR4 7TJ, United Kingdom

**ABSTRACT**

This paper develops, in precise quantum electrodynamic terms, photonic attributes of the "optical chirality density", one of several measures long known to be conserved quantities for a vacuum electromagnetic field. The analysis lends insights into some recent interpretations of chiroptical experiments, in which this measure, and an associated chirality flux, have been treated as representing physically distinctive "superchiral" phenomena. In the fully quantized formalism the chirality density is promoted to operator status, whose exploration with reference to an arbitrary polarization basis reveals relationships to optical angular momentum and helicity operators. Analyzing multi-mode beams with complex wave-front structures, notably Laguerre-Gaussian modes, affords a deeper understanding of the interplay between optical chirality and optical angular momentum. By developing theory with due cognizance of the photonic character of light, it emerges that only the spin angular momentum of light is engaged in such observations. Furthermore, it is shown that these prominent measures of the helicity of chiral electromagnetic radiation have a common basis, in differences between the populations of optical modes associated with angular momenta of opposite sign. Using a calculation of the rate of circular dichroism as an example, with coherent states to model the electromagnetic field, it is discovered that two terms contribute to the differential effect. The primary contribution relates to the difference in left- and right- handed photon populations; the only other contribution, which displays a sinusoidal distance-dependence, corresponding to the claim of nodal enhancements, is connected with the quantum photon number-phase uncertainty relation. From the full analysis, it appears that the term "superchiral" can be considered redundant.

# 1. Introduction

Recently, considerable interest has been aroused by a rediscovered measure of helicity in optical radiation – commonly termed "optical chirality density". This time-even, parity-odd pseudoscalar has been shown to be key in determining, for example, the differential absorption of circularly polarized light in small chiral enantiomers (molecules of opposite handedness), along with a host of other optically active processes. Specifically, when considering the regular dipolar mechanisms for optical excitation, any difference in rates of absorption proves to be proportional to a product of the local optical chirality – given by the spatially integrated optical chirality density – and another pseudoscalar characterizing the inherent chirality of the material. Originally termed the "Lipkin zilch", the optical chirality measures were found to be distinct from stress-energy and were initially dismissed as having no ready physical interpretation, but were later associated with conservation of polarization of the electromagnetic field [1, 2]. Recent work by Bliokh and Nori has uncovered close connections between the optical chirality density and such measures as polarization helicity and energy density [3].

Many biomacromolecules are composed of intrinsically chiral molecular sub-units, such as sugars and amino acids, and molecular chirality is therefore often used for detection and characterization purposes, through the deployment of chiroptical spectroscopic techniques such as circular dichroism (CD), optical rotation and Raman optical activity. For this reason, recent claims of generating and detecting "superchiral" light, whose electromagnetic fields have a chiral dissymmetry greater than that of pure circularly polarized light, have been met with enthusiasm [4]. Tang and Cohen recently assembled an experiment to detect differential fluorescence near to one node of a circularly polarized electromagnetic standing wave, a region postulated to be "superchiral". They irradiated a sample consisting of an achiral control layer and a neighboring chiral layer, each 10 nm in thickness, with a green (543 nm) laser beam, the layers being set at a tilt to ensure that part of the sample was near to a node. On the side of the sample opposing the laser was a partial mirror reflecting the radiation, the superposition of the counter-propagating beams generating a standing wave [5, 6]. Under such conditions, observations of differential absorption one or two orders of magnitude larger than expected were interpreted as a manifestation of superchirality.

Kadodwala et al. recently provided experimental results that were similarly interpreted as verifying the capability of the optical chirality to identify regions of uncommonly enhanced chiral dissymmetry. They used planar chiral metamaterials (PCMs) comprising left- and right- handed gold gammadions of length 400 nm and thickness 100 nm, with a 5-nm chromium adhesion layer, deposited on a glass substrate with a periodicity of 800 nm. Using UV-visible CD spectroscopy, the optical properties of the PCMs were probed under various liquid layers, and resonances in the CD spectra, associated with the excitation of localized surface plasmon resonances, were observed. Furthermore, it was noted that there exist regions in which the effective chirality of the field appears "one to two orders of magnitude larger than expected for circularly polarized plane waves" [7]. This is in agreement with the well known surface plasmonic amplification of local electric fields in systems fabricated with a metal substrate, in which chiroptical (as well as conventional optical) response will exhibit much larger than usual effects [8]. In support of their conclusions, Kadodwala et al. exhibited the results of calculating both the electric field strengths and the corresponding value of the optical chirality around the PCM [9].

One therefore has to ask what property there could be, in supposedly "superchiral light", that could support a higher than usual degree of optical chirality, capable of engaging with electronic transitions that can exhibit a discriminatory response to opposite helicities. Because of the deep fundamental link between optical angular momentum and chirality, the only candidate is optical orbital angular momentum – the counterpart spin component having a well-known connection with the circularity of polarization. In the following work we prove that the value of the optical chirality is related only to the helicity and spin angular momentum of the electromagnetic field. Any "superchiral" behavior would require the additional engagement of the field's orbital angular momentum [10-12]; however, we shall show that photon spin alone can engage in circular dichroic effects; moreover we explicitly demonstrate that orbital angular momentum cannot be responsible for the reported effects.

## 2. Measures of chirality in optical radiation

In order to more fully understand the reported phenomenon it is necessary to adopt a completely quantized representation of the electromagnetic field. In this quantum optical formulation the fields and related variables are promoted to quantum operators. In particular it is appropriate to consider in detail the optical angular momentum operator, in order to clarify the relationship between it and other measure of electromagnetic helicity.

In general, the angular momentum **J** of the electromagnetic field can be defined as [13]:

$$\mathbf{J} = \varepsilon_0 \int d^3\mathbf{r}\, \mathbf{r} \times (\mathbf{E} \times \mathbf{B}), \qquad (1)$$

where $\varepsilon_0$ is the vacuum permittivity, **r** is the position vector and **E**, **B**, respectively, are the electric and magnetic induction fields, implicitly evaluated at that position. It is well known that this total angular momentum operator can be recast as the sum of the following terms:

$$\mathbf{L} = \varepsilon_0 \hat{r}_i \int d^3\mathbf{r}\, E_l (\mathbf{r} \times \nabla)_i \cdot A_l, \tag{2}$$

$$\mathbf{S} = \varepsilon_0 \hat{r}_i \int d^3\mathbf{r}\, (\mathbf{E} \times \mathbf{A})_i, \tag{3}$$

**L** signifying the orbital angular momentum for the field, and **S** the spin angular momentum; **A** is the electromagnetic vector potential, and the Einstein summation convention is used. Indeed, the spin angular momentum need not be derived from equation (1) as it is known that it arises separately from Maxwell's equations, when intrinsic torque densities are included. It is now shown that a mode analysis on the spin angular momentum (SAM) operator defined in equation (3) permits its expression in terms of photon creation and annihilation operators $a^{(\eta)}(\mathbf{k})$ and $a^{\dagger(\eta)}(\mathbf{k})$. The operator for the electromagnetic vector potential is given by:

$$\mathbf{A} = \sum_{\mathbf{k},\eta} \left( \frac{\hbar}{2\varepsilon_0 ckV} \right)^{\frac{1}{2}} \left\{ \mathbf{e}^{(\eta)}(\mathbf{k}) a^{(\eta)}(\mathbf{k}) e^{i(\mathbf{k}\cdot\mathbf{r})} + \overline{\mathbf{e}}^{(\eta)}(\mathbf{k}) a^{\dagger(\eta)}(\mathbf{k}) e^{-i(\mathbf{k}\cdot\mathbf{r})} \right\}, \tag{4}$$

where $\hbar$ is the reduced Planck constant, $V$ is the quantization volume, and $\mathbf{e}^{(\eta)}(\mathbf{k})$ is the polarization vector for a mode with polarization $\eta$ and wave-vector **k**; the right-hand term in (4) represents the Hermitian conjugate of the term on the left [14]. The magnetic and electric fields then emerge in operator form through the prescription:

$$\mathbf{B} = \nabla \times \mathbf{A}; \quad \mathbf{E} = -\partial \mathbf{A}/\partial t \tag{5}$$

allowing completion of the mode analysis on equation (3); the final result is then delivered as:

$$\mathbf{S} = \hbar \sum_{\mathbf{k}} \left\{ \hat{N}^{(L)}(\mathbf{k}) - \hat{N}^{(R)}(\mathbf{k}) \right\} \hat{\mathbf{k}}, \tag{6}$$

where $\hat{N}^{(\eta)}(\mathbf{k}) = a^{\dagger(\eta)}(\mathbf{k}) a^{(\eta)}(\mathbf{k})$ is the photon number operator, and $L/R$ correspond to left and right circular polarizations respectively. As might be anticipated, the spin angular momentum operator depends solely on the disparity of left- and right- handed photon populations. In the detailed mode analysis, it will be necessary to enact a sum over polarization states through the embedding of equation (4); here we note that the circularly polarized basis can be expressed in terms of the following unit vectors:

$$\mathbf{e}^{(L)}(\mathbf{k}) = \frac{1}{\sqrt{2}} \left( \hat{\mathbf{i}} + i\hat{\mathbf{j}} \right); \quad \mathbf{e}^{(R)}(\mathbf{k}) = \frac{1}{\sqrt{2}} \left( \hat{\mathbf{i}} - i\hat{\mathbf{j}} \right), \tag{7}$$

where $\hat{\mathbf{i}}$ and $\hat{\mathbf{j}}$ are the Cartesian unit vectors, such that $(\hat{\mathbf{i}}, \hat{\mathbf{j}}, \hat{\mathbf{k}})$ comprise a right-handed orthogonal group.

The aim is to now show that this direct dependence on the difference between the number operators for left- and right- handed modes appears in a variety of electromagnetic helical measures. We proceed by first analyzing the helicity of the free electromagnetic field, defined as:

$$\kappa = \int d^3\mathbf{r}\, \mathbf{A} \cdot \mathbf{B} \tag{8}$$

[15, 16]. Following the prescription of mode analysis used with the SAM operator, the helicity operator emerges as:

$$\kappa = \frac{\hbar}{\varepsilon_0 c} \sum_k \left( \hat{N}^{(L)}(\mathbf{k}) - \hat{N}^{(R)}(\mathbf{k}) \right). \tag{9}$$

Notably the result is again dependent on the difference between the number operators for the respective modes, characterized by wave-vector **k**.

Our main focus is now on a similar mode analysis on the optical chirality density, where it will be instructive to consider in more detail the symmetry properties of this measure. First, the optical chirality density is defined as

$$\chi = \frac{\varepsilon_0}{2} \mathbf{E} \cdot \nabla \times \mathbf{E} + \frac{1}{2\mu_0} \mathbf{B} \cdot \nabla \times \mathbf{B}, \tag{10}$$

where $\mu_0$ is the vacuum permeability. In the ensuing quantum electrodynamic (QED) representation, $\chi \equiv \chi(\mathbf{r},t)$ is to be regarded as an operator on the radiation states. In terms of fundamental symmetries, the matrix elements of $\chi$ in a radiation state basis are pseudoscalars, odd with respect to the operator for space inversion (or parity), $\mathcal{P}$, but even under the time reversal operation, $\mathcal{T}$ [17]. Using a mode expansion on equation (10), similar to that enacted upon the spin and helicity operators, proves to deliver a result whose expectation value for a particular optical state is also dependent on the difference between the left- and right- handed photon populations. To further comprehend the physical picture we note that χ satisfies the following continuity equation:

$$\frac{\partial \chi}{\partial t} + \nabla \cdot \boldsymbol{\varphi} = 0. \tag{11}$$

with respect to the corresponding flux of χ;

$$\varphi = \frac{\varepsilon_0 c^2}{2}\left[\mathbf{E}\times(\nabla\times\mathbf{B}) - \mathbf{B}\times(\nabla\times\mathbf{E})\right], \tag{12}$$

hereafter called the "optical chirality flow". In a fully closed system, chirality is a conserved quantity; this is a logical consequence of $\mathcal{CPT}$ invariance. As regards the expectation values for these two measures in a state comprising a different number of left- and right-handed photons, the chirality is also accordingly conserved, in the absence of any sinks (or sources) – the latter signifying physical processes of photon absorption (or emission). Indeed, any conservation law such as that presented in equation (11) can be written as $\partial^\mu \varphi_\mu = 0$, for $\mu = 0, 1, 2, 3$, where the zeroth index represents the time component and the further indices represent spatial components; as such, the optical chirality and chirality flow together represent components of a 4-vector $(c\chi, \varphi)$ in Minkowski space [2, 18].

We have reached the optimal point in the analysis to undertake a mode expansion of the expression for the optical chirality density presented in equation (10). First we note in passing that, as is the case with the helicity and spin angular momentum operators, use of a linearly polarized light basis for the polarization sum in the optical chirality density gives a null result, as can be expected for light fields with no angular momentum. When using the circularly polarized light basis, shown in equation (7), the calculations produce the following equation:

$$\int d^3\mathbf{r}\, \chi = \sum_\mathbf{k} \hbar c k^2 \left\{ \hat{N}^{(L)}(\mathbf{k}) - \hat{N}^{(R)}(\mathbf{k}) \right\}, \tag{13}$$

where it can be seen that the optical chirality density is dependent on the same difference in left and right photon populations as the previously analyzed measures on the electromagnetic field. Furthermore, when considering a monochromatic (not necessarily parallel) beam, of circular frequency $\omega = ck$, we obtain the elementary result:

$$\int d^3\mathbf{r}\, \chi = \varepsilon_0 \omega^2 \int d^3\mathbf{r}\, \mathbf{A}\cdot\mathbf{B}. \tag{14}$$

For a generalization of this analysis it is appropriate at this juncture to allow a relaxation of two previous assumptions. First, we accommodate a degree of freedom in the choice of basis polarization vectors, allowing a generalization of the result for circularly polarized light. Any acceptable basis of states has to satisfies the necessary orthogonality condition, $\mathbf{e}_n \cdot \bar{\mathbf{e}}_m = \delta_{nm}$, the polarization pair corresponding to diametrically opposite points on the Poincaré sphere [19]. An arbitrary polarization vector $\mathbf{e}_1$, characterized by angular coordinates $\theta$ and $\phi$, and its counterpart basis vector $\mathbf{e}_2$, are generated according to the following prescription:

$$\left.\begin{aligned}\mathbf{e}_1 &= \sin\theta\hat{\mathbf{i}} + e^{i\phi}\cos\theta\hat{\mathbf{j}}\\ \mathbf{e}_2 &= \cos\theta\hat{\mathbf{i}} - e^{i\phi}\sin\theta\hat{\mathbf{j}}\end{aligned}\right\}. \qquad (15)$$

Secondly, to address the possible involvement of orbital angular momentum it is furthermore expedient to consider, as a representative example, Laguerre-Gaussian (LG) modes, these being prototypical examples of beams bearing orbital angular momentum [20, 21]. As has recently been shown, the field structures of such beams can also be represented by an extension of the Poincaré sphere [22]. In the usual paraxial approximation, the magnetic and transverse electric field vectors are determined by the LG vector potential given by;

$$\mathbf{A} = \sum_{\mathbf{k},\eta,l,p}\left(\frac{\hbar}{2\varepsilon_0 ckV}\right)^{\frac{1}{2}}\left\{\mathbf{e}_{l,p}^{(\eta)}(\mathbf{k})a^{(\eta)}(\mathbf{k})f_{l,p}(r)e^{ikz-il\varphi} + \bar{\mathbf{e}}_{l,p}^{(\eta)}(\mathbf{k})a^{\dagger(\eta)}(\mathbf{k})f_{l,p}(r)e^{-ikz+il\varphi}\right\} \qquad (16)$$

where $f_{l,p}(r)$ represents the radial distribution of the LG mode with radial number $p$ and azimuthal index $l$. Once again, the **E** and the **B** fields are obtained by the prescription noted in equation (5). With the generalized polarization basis (15), the result for the integrated chirality density duly emerges as:

$$\int d^3\mathbf{r}\,\chi = \hbar c\sin 2\theta\sin\varphi\sum_{\mathbf{k}}k^2\left\{\hat{N}^{(1)}(\mathbf{k}) - \hat{N}^{(2)}(\mathbf{k})\right\}, \qquad (17)$$

where the superscripts 1 and 2 correspond to the arbitrarily chosen basis vectors. Recognizing the orthogonality of the radial distribution function, the $e^{-il\varphi}$ and $f_{l,p}(r)$ factors included in the expression for the LG mode disappear from the calculation on implementing the normalization condition on the quantization volume, and the result (17) is therefore independent of any factors pertaining to orbital angular momentum.

The significance in the result is this: the optical chirality density is *not* a measure of orbital angular momentum, only spin. It follows that any field whose chiral character is fully defined by the optical chirality measure cannot possibly display "superchiral" behavior; the maximum value the expression (17) can take is when the radiation is circularly polarized (wherein the sinusoidal factors have the value of unity), and when the population of the second basis state is zero. When the generalized polarization representation is used in the plane mode analysis, as performed earlier with circularly polarized light, the emergent result is identical to that expressed in equation (17). Furthermore, as the determination of equation (17) does not explicitly depend on the specific LG structure of the radial distribution function, only the associated orthogonality condition, it is evident that other well known orthogonal solutions to the paraxial wave equation will display identical expressions for the spatially integrated optical chirality density. For example, Airy, Bessel, Hermite-Gauss, and Mathieu beams, all have expressions for the optical chirality operator proportional to the difference of the number operators in the polarization basis. For completeness we note that the mode expansion for the optical chirality flux is given by

$$\varphi = \hbar\varepsilon_0 c^2 \sin 2\theta \sin\varphi \sum_{\mathbf{k}} k^2 \left\{ \hat{N}^{(1)}(\mathbf{k}) - \hat{N}^{(2)}(\mathbf{k}) \right\} \hat{\mathbf{k}}, \qquad (18)$$

which, for a monochromatic (not necessarily parallel) beam, is equivalent to $\varepsilon_0 c^2 k^2$ times the expression for the SAM operator. This equivalence, taken with the conservation equation governing the optical chirality and flux (11), signifies that for a monochromatic beam, electromagnetic helicity is conserved with respect its flux, the spin angular momentum – exhibiting the fact that the helicity is the projection of the spin onto the direction of the momentum.

## 3. Framework for connectivity of molecular and optical chirality

We now introduce a comprehensive framework for capturing the mechanism by means of which chirality in a radiation field can engage with chirality in matter, and revealing the origin of the symmetry principles involved. For generality, we begin with a formulation that is applicable to spectroscopic processes of an arbitrary order of optical nonlinearity, using a multipolar representation of the coupling [23]. Each photon interaction entails an interaction Hamiltonian that comprises linear couplings of the molecular polarization field **p**(**r**) (accommodating all electric multipoles) with the transverse electric field $\mathbf{e}^\perp(\mathbf{r})$ of the radiation, and couplings of the molecular magnetization field **m**(**r**) (constituent magnetic multipoles) with the magnetic induction field **b**(**r**):

$$H_{\text{int}} = -\int d^3\mathbf{r} \left[ \mathbf{p}(\mathbf{r}) \cdot \mathbf{e}^\perp(\mathbf{r}) + \mathbf{m}(\mathbf{r}) \cdot \mathbf{b}(\mathbf{r}) \right] \qquad (19)$$

Equation (19) excludes a diamagnetization term, associated with variations of current density, that is quadratic in the magnetic field: however the associated couplings are smaller by an order of magnitude or more than those that relate to the two linear terms exhibited in (19), and there is no mechanism for their involvement in chiral phenomena. The usual Taylor series expansion of the linear terms separates the coupling into multipolar orders, whose leading contributions are E1, E2 and M1, cast in terms of quantum operators for the electric dipole **μ**, electric quadrupole **Q** and magnetic dipole **m**; couplings involving the latter derive from the same level of expansion in the minimal coupling representation, but they are smaller than the µ term by a factor of the order of the fine structure constant [14].

Let us consider a process in which *m* photons are involved in each fundamental interaction with a material centre such as an atom or molecule. Under the normal conditions that support the determination of a rate from time-dependent perturbation theory, the key observables are determined from the square modulus of a scalar matrix element $M_{fi}^{(m)}$ that can be cast as follows [24];

$$M_{fi}^{(m)} = \sum_{e=0}^{m} \sum_{b=m-e}^{m} K_{e;b}^{(m)} \mathbf{S}_{e;b;m-e-b}^{(e+b+2q)} \otimes \mathbf{T}_{e;b;m-e-b}^{(e+b+2q)} . \qquad (20)$$

Here, the expression on the right comprises a sum of terms, each expressed in the form of an inner tensor product between radiation tensors $\mathbf{S}_{e;b;m-e-b}^{(e+b+2q)}$ and corresponding-rank molecular tensors $\mathbf{T}_{e;b;m-e-b}^{(e+b+2q)}$. Each such tensor is distinguished by labels (*e*, *b*, *q*) corresponding to the number of E1, M1, and E2 interactions, respectively; the sum of these equals the number of photon interactions involved in the process, *m* = *e* + *b* + *q*. The radiation tensor $\mathbf{S}^{(e+b+2q)} \equiv S_{i_1 i_2 ... i_r}$ duly comprises an outer product of electric and magnetic induction field components, and electric field gradients, resulting from a product of transition integrals (Dirac brackets) for the radiation tensor. Notice that, because these transition integrals are implemented for radiation states that represent the conditions of a specific experimental interaction, they can exhibit a lower symmetry than that of the field operators they contain. For example, one matrix element of **S** might exhibit the helical symmetry properties of a particular circular polarization, if only that polarization is present in the radiation (whereas the mode expansion would have both helicities equally represented, and therefore have no overall helical character). The corresponding molecular tensor $\mathbf{T}^{(e+b+2q)} = T_{i_1 i_2 ... i_r}$, can be written in a form entailing a product of *m* molecular transition integrals. For any such transition integral not to vanish identically, the product of the symmetry representations for the two states that it connects must span one or more of the irreducible representations under which components of the corresponding transition moment itself transforms, under the full set of symmetry operations determined by the molecular point group [25]. Mapping the irreducible representations of the full three-dimensional rotation-inversion group O(3) onto a lower symmetry in many cases permits transitions to occur between states of more than one symmetry class, a necessary condition for exhibiting molecular chirality.

We can now interrogate equation (20) for its fundamental symmetry behavior. The matrix element is a scalar with the dimensions of energy, obviously invariant with respect to space or time inversion. It is important to note that although the constants, $K_{e;b}^{(m)}$, have physical dimensions determined by $e$ and $b$, they also are invariant with respect to space and time inversion; all the dynamical symmetries are accommodated within the radiation and molecular tensors. Clearly, the parity signatures of each corresponding **S** and **T** tensor have to be identical, both with respect to the operations of space inversion $\mathcal{P}$ and time reversal $\mathcal{T}$. In fact, these signatures are $(-1)^e$ and $(-1)^b$, respectively, as determined by the space-odd, time-even character of the electric field, and the space-even, time-odd character of the magnetic field. Notice that the possible involvement of an electric quadrupole plays no part in this determination, because of its even parity under both $\mathcal{P}$ and $\mathcal{T}$.

When the rate is evaluated from the square modulus of the matrix element (20), 'diagonal' terms do not produce any distinctive behavior with regard to fundamental symmetry. For example a quadratic dependence on any radiation tensor **S** is necessarily of even parity with respect to both space and time; the same is true for the corresponding **T**. Thus, terms that are quadratically dependent on any specific molecular tensor (such as, in the simplest case, one electric dipole transition moment) will be invariant under a space inversion that physically corresponds to a substitution of the opposite enantiometric form, in the case of a chiral molecule. The same logic applies to the radiation tensor, with regard to an inversion that signifies a change to circular polarization of the opposite handedness.

The quantum interferences between terms with different symmetry character, in the overall rate equation, are therefore of primary interest. The distinctive feature of a chiral centre is that, included in these interference terms are products that are odd in spatial parity. For example one specific matrix element contribution might be written as $\mathbf{S}^{(r)} \otimes \mathbf{T}^{(r)}$ and another as $\mathbf{S}^{(t)} \otimes \mathbf{T}^{(t)}$; the products $\mathbf{S}^{(r)} \mathbf{S}^{(t)}$ and $\mathbf{T}^{(r)} \mathbf{T}^{(t)}$ may have a net parity of -1 with respect to either $\mathcal{P}$ or $\mathcal{T}$, notwithstanding the fact that their product is necessarily of even parity. It suffices to focus on the space parity, for which the conditions supporting a non-zero optical chirality density $\chi$ of equation (10), itself a pseudoscalar, clearly also support the odd-parity instances of $\mathbf{S}^{(r)} \mathbf{S}^{(t)}$. As we have seen, this necessitates a radiation field whose arbitrary basis modes – according to the prescription given in (15) – are disproportionately populated, most notably a non-zero difference in circularly polarized photon populations.

**4. Circular dichroism: a test case**

The conjecture that the optical chirality metrics might signify more comprehensive measures of differential chiral response can be tested by considering circular dichroism, a classic example of an interaction whose dependence can be calculated by symmetry methods. Since the initial and final molecular states differ, the process is incoherent, obviating any interference between quantum amplitudes associated with different molecules [26]. With the engagement of only one photon, the Fermi rate for the absorption is given by:

$$\Gamma \sim N_\xi \left\langle \left| M_{fi}^{\xi} \right|^2 \right\rangle = N_\xi \left\langle \left| M_{fi}^{\xi(\text{E1})} + M_{fi}^{\xi(\text{M1})} + \ldots \right|^2 \right\rangle$$
$$= N_\xi \left[ \left\langle \left| M_{fi}^{\xi(\text{E1})} \right|^2 \right\rangle + 2\Re e : \left\langle M_{fi}^{\xi(\text{E1})} \overline{M_{fi}^{\xi(\text{M1})}} \right\rangle + \left\langle \left| M_{fi}^{\xi(\text{M1})} \right|^2 \right\rangle + \ldots \right], \qquad (21)$$

where $N_\xi$ is the number of molecules, $\Re e$: denotes the real part, and angular brackets signify rotational averaging, required in the case of a fluid sample. The leading $(\text{E1})^2$ term and the $(\text{M1})^2$ term in (21) deliver identical results for either left- or right-handed circularly polarized radiation, whereas the E1-M1 interference changes sign if the helicity of the radiation (or molecular handedness) is reversed. In such chiroptical phenomena, the largest contributions generally come from the E1-M1 and E1-E2 interactions, but the latter contribution vanishes when an isotropic rotational average is taken. Thus, the emergent rate differential for circular dichroism emerges as:

$$\left\langle \Gamma_\text{L} \right\rangle - \left\langle \Gamma_\text{R} \right\rangle \sim 4\Re e : N_\xi \left\langle M_{fi}^{\xi}(\text{E1}) \overline{M_{fi}^{\xi}(\text{M1})} \right\rangle. \qquad (22)$$

It is readily seen from equation (22) that handedness is apparent in two respects; using the parity operation on just the molecular multipoles (not the radiation) – corresponding to a change of enantiomer – invokes a sign change in the interference term. Similarly, reversing the handedness of the radiation, whilst leaving the molecular multipoles unchanged, again results in the interference term changing sign.

*4.1 Single-beam case*

We now tackle the explicit quantum electrodynamical calculation of the differential absorption associated with an electronic transition $0 \rightarrow \beta$ in a chiral molecule, characterized by electric and magnetic transition dipole moments, $\boldsymbol{\mu}^{\beta 0}$ and $\mathbf{m}^{\beta 0}$, respectively. The difference in Fermi rates of single-beam absorption in a system of $N_\xi$ left- and right- handed enantiomers, denoted by + and − respectively, gives:

$$\left\langle \Gamma^{(+)} \right\rangle - \left\langle \Gamma^{(-)} \right\rangle \sim \Im m : N_\xi \boldsymbol{\mu}^{\beta 0(+)} \cdot \mathbf{m}^{\beta 0(+)}$$
$$\times \sum_{\mathbf{k}} k \left\langle n^{(L)}(\mathbf{k}), n^{(R)}(\mathbf{k}) \left| \hat{N}^{(L)}(\mathbf{k}) - \hat{N}^{(R)}(\mathbf{k}) \right| n^{(L)}(\mathbf{k}), n^{(R)}(\mathbf{k}) \right\rangle, \qquad (23)$$

where the input radiation mode of wave-vector $\mathbf{k}$ comprises $n^{(L)}$ and $n^{(R)}$ photons of left and right-handed circularity, respectively, and $\Im m$: indicates the imaginary part of the expression. Equivalently:

$$\left\langle \Gamma^{(+)} \right\rangle - \left\langle \Gamma^{(-)} \right\rangle \sim \Im m : N_\xi \left( \mathbf{\mu}^{\beta 0(+)} \cdot \mathbf{m}^{\beta 0(+)} \right) \int d^3 \mathbf{r} \, \chi \, . \tag{24}$$

Here it is explicitly evident that the difference in the rates of absorbing left- and right-handed light is given by a product of the spatially integrated optical chirality with the pseudoscalar $\mathbf{\mu}^{\beta 0(+)} \cdot \mathbf{m}^{\beta 0(+)}$, the latter representing the inherent chirality of the material.

*4.2 Counterpropagating beams*

Considering that the experiments of most interest are performed with counter-propagating beams, it is expedient for further calculations to reflect this fact in the guise of a superposition state. To properly exhibit the important phase properties associated with the corresponding electromagnetic field, both the incident and reflected beams are best described by a coherent radiation state – one that has minimum quantum uncertainty and is most like a classical wave [27]. Photon number (Fock) states are not eigenstates of the annihilation operator, and so the absorption of light from different modes could not in fact give the required interference. Specified by a complex variable $\alpha$, the coherent state $\left| \alpha(\mathbf{k}, \eta) \right\rangle$ is an eigenstate of the photon annihilation operator:

$$\hat{a}^{(\eta)}(\mathbf{k}) \left| \alpha(\mathbf{k}, \eta) \right\rangle = \alpha(\mathbf{k}, \eta) \left| \alpha(\mathbf{k}, \eta) \right\rangle, \tag{25}$$

and the probability of a given number of photons being measured in a coherent state follows a Poison distribution, thus:

$$\overline{\hat{N}^{(\eta)}(\mathbf{k})} = |\alpha|^2 \, ; \, \Delta n = |\alpha| \, . \tag{26}$$

Here the overbar represents the expectation value, and $\Delta n$ the uncertainty, for the photon population. We are now in a position to characterize the initial and final states of the system:

$$\left| \text{mol} \right\rangle \left| \text{rad} \right\rangle = \left| \psi_{0/\beta} \right\rangle \left| \alpha \left( \mathbf{k}, e^{(L)}(\mathbf{k}) \right), \alpha' \left( \mathbf{k}', e'^{(R)}(\mathbf{k}') \right) \right\rangle, \tag{27}$$

where the molecular state is characterized by the wavefunction for the molecule (0/β for the ground and excited state, respectively) and the radiation comprises a superposition of the coherent states α and α', with wave-vectors **k** and **k′** = −**k**, respectively. In absorption the leading term in the rate corresponds to the E1$^2$ term, as such the principal contribution to the probability amplitude is:

$$\Gamma^{(\pm)} \sim \Re e : N_\xi e_i^{(L)}(\mathbf{k}) e_j^{(R)}(\mathbf{k}) \mu_i^{\beta 0} \mu_j^{\beta 0} \left\{ \overline{n}_L + \overline{n}_R' - 2 \Delta n_L \Delta n_R' e^{2ikz} \right\}, \tag{28}$$

where $z$ is the position relative to the mirror and the expression remains unchanged when swapping enantiomer. If the input and reflected beams have similar α values, then the coherent state properties (26) imply that the corresponding photon populations are similar and hence the rate, given in equation (28), has a position dependence on $1-\cos(2kz)$. Therefore, nodes – where the expression vanishes – appear at $z = n\pi/k, n \in \mathbb{Z}$. However, as before, the largest contribution to the *differential* absorption is the E1-M1 interference term, from which the difference between the absorption rates emerges as:

$$\Gamma^{(+)} - \Gamma^{(-)} \sim \Re e : N_\xi e_i^{(L)}(\mathbf{k}) e_j^{(R)}(\mathbf{k}) \left\{ \left( i\mu_i^{\beta 0(+)} m_j^{\beta 0(+)} + im_i^{\beta 0(+)} \mu_j^{\beta 0(+)} \right) \left( \overline{n}_R' - \overline{n}_L \right) \right.$$
$$\left. + \left( i\mu_i^{\beta 0(+)} m_j^{\beta 0(+)} - im_i^{\beta 0(+)} \mu_j^{\beta 0(+)} \right) \left( 2\Delta n_L \Delta n_R' e^{2ikz} \right) \right\}. \qquad (29)$$

One part of the result is, as in single-beam case, dependent on the difference in the number of left- and right-handed photons – here signified by the input and reflected photon populations – through the term $\left( \overline{n}_R' - \overline{n}_L \right)$. However, the above result also contains an interference term that is dependent on the position $z$, relative to the mirror. Significantly, this term vanishes on taking an isotropic rotational average – signifying the conditions of a fluid for which the normal measures of circular dichroism are defined. Nonetheless under the same rotationally averaged conditions, the first term of (29) generates the pseudoscalar $\boldsymbol{\mu}^{\beta 0(+)} \cdot \mathbf{m}^{\beta 0(+)}$, which can only be non-zero for chiral molecules.

We can verify that the interference term also satisfies the necessary symmetry constraints, changing sign under parity inversion (equivalent to swapping the handedness of the input radiation $L \leftrightarrow R$). In the framework of quantum mechanics the magnetic dipole operator is given

$$\mathbf{m} = -\frac{i\hbar\rho}{2} \int d^3\mathbf{r} \, \mathbf{r} \times \nabla. \qquad (30)$$

where ρ is the charge density. Accordingly, choosing the electric dipole moment to be real requires the magnetic dipole moment to be imaginary; thus we can write $i\mathbf{m} = \mathbf{M}$, where $\mathbf{M}$ is real. Casting the vector indices in (29) in terms of specific Cartesian components using equation (7), we thus have:

$$\Gamma^{(+)} - \Gamma^{(-)} \sim N_\xi \left\{ \left[ \mu_x^{\beta 0(+)} M_x^{\beta 0(+)} + \mu_y^{\beta 0(+)} M_y^{\beta 0(+)} \right] \left( \overline{n}_L - \overline{n}_R' \right) \right.$$
$$\left. + 4 \left[ \mu_x^{\beta 0(+)} M_y^{\beta 0(+)} - \mu_y^{\beta 0(+)} M_x^{\beta 0(+)} \right] \Delta n_L \Delta n_R' \cos(2kz) \right\}, \qquad (31)$$

from which it is clear that the whole expression duly changes sign under parity inversion.

More generally, we note that the position-determined conditions under which the interference term vanishes, namely for $z = (2n-1)\pi/4k, n \in \mathbb{Z}$, signify positions at which the electric vector of one beam is parallel (or anti-parallel) to the magnetic vector of the other. Molecules interacting with these two field components therefore experience an electromagnetic influence that has the symmetry of a plane containing those vectors and the propagation direction; these do not span the three-dimensional vector space $\mathbb{R}^3$ – in other words there is, at such positions, no three-dimensional basis for resolving chirality and generating a circular differential.

The first term in (31), through its dependence on the difference in mean left- and right-handed photon populations, relates directly to the measures of helicity for the electromagnetic field, most notably the optical chirality $\chi$ (17). The term with sinusoidal distance-dependence notably exhibits a connection with phase-photon number uncertainty relation through the product of $\Delta n$'s – defined in equation (26) – and therefore displays shot noise, a feature associated with the Poisson distribution of photons in a coherent state [28]. Significantly, this term corresponds to the claim of nodal enhancements in the experiments devised by Tang and Cohen [5, 6].

## 5. Conclusion

We have shown that CD and related chiroptical phenomena respond only to the polarization state, and that the extent of such effects in beams with precise number states cannot exceed that delivered by a circularly polarized beam. Furthermore, it has been shown that a precise QED representation of recently performed experiments shows circular differential results due to E1-M1 interference with two terms: one position-independent term that is characteristically dependent on differences of photon populations in a pair of orthogonal basis states, and another with a sinusoidal dependence on position. However, the nodal positions for the latter term do not coincide with the positions at which zeros occur in the normally dominant $E1^2$ contribution to the absorption rate. The experiment performed by Tang and Cohen was initially regarded as indicating that regions near to an electromagnetic node experience "superchiral" fields – with a dissymmetry factor greater than that for purely circularly polarized light [5]. The above analysis shows that although there are certainly nodal enhancements (or attenuations) to the differential absorption of circularly polarized light, the effect is a result of the beam superposition.

Although it is certainly possible to generate optical states of still more highly chiral character, consistent with Maxwell's equations, through the engagement of orbital angular momentum, the additional contributions to optical chirality cannot be measured by spectroscopic means [29, 30]. When such effects become especially prominent at metal surfaces, through the generation of surface plasmon optical vortices [8, 31], they are manifest in mechanical rather than chiroptical effects. It is well known that planar surface structures, whilst not intrinsically chiral, can support chiral reponse when coupled with a physical stimulus whose sense of direction has a component perpendicular to the plane. The studies reported by Kadodwala et al. are consistent with this setup; radiation impinging on the PCM structures will accordingly have the capacity to exhibit circular differential response. In particular, molecules adsorbed on such structures will certainly exhibit circular dichroism at wavelengths where they absorb. However, it can be misleading to apply to surface adsorbed species, formulae designed to quantify the extent of CD in solution media. It is not surprising that truly normalized measures of differential response, such as the dissymmetry factor, should be larger in the system studied than for a corresponding solution.

**Acknowledgment**

The authors would like to thank EPSRC for funding this research.

* E-mail: david.andrews@physics.org